\documentclass[conference,a4paper]{IEEEtran}
\IEEEoverridecommandlockouts
\usepackage{cite}
\usepackage{amsmath,amssymb,amsfonts}
\usepackage{algorithmic}
\usepackage{graphicx}
\usepackage{textcomp}
\usepackage{xcolor}
\def\BibTeX{{\rm B\kern-.05em{\sc i\kern-.025em b}\kern-.08em
    T\kern-.1667em\lower.7ex\hbox{E}\kern-.125emX}}

\newcommand{\uptxt}[1]{^{\mathrm{#1}}}
\usepackage{subfig}
\usepackage{nomencl}
\makenomenclature
\usepackage{etoolbox}
\renewcommand\nomgroup[1]{%
  \item[\bfseries
  \ifstrequal{#1}{A}{Sets and Indices}{
  \ifstrequal{#1}{B}{Parameters}{
  \ifstrequal{#1}{C}{Variables}{}}}
]}

\begin{document}

\title{Optimal Operation of a Building with Electricity-Heat Networks and Seasonal Storage
}

\author{
\IEEEauthorblockN{%
        Eléa Prat, 
        Pierre Pinson, 
        Richard M. Lusby,
        Riwal Plougonven,
        Jordi Badosa,
        Philippe Drobinski
        }
\thanks{E. Prat (e-mail: emapr@dtu.dk) and R. Lusby are with the Technical University of Denmark, Kgs. Lyngby, Denmark.}
\thanks{P. Pinson has primary affiliation with Imperial College London, UK. He has additional affiliations with Halfspace (DK), the Technical University of Denmark and Aarhus University (CoRE).}
\thanks{R. Plougonven, J. Badosa and P. Drobinski are with LMD-IPSL, Ecole Polytechnique, Institut Polytechnique de Paris, ENS, PSL Research University, Sorbonne Université, CNRS, 91120 Palaiseau, France.}
\thanks{This work contributes to the Energy4Climate Interdisciplinary Center (E4C) of IP Paris and Ecole des Ponts ParisTech, supported by 3rd Programme d’Investissements d’Avenir [ANR-18-EUR-0006-02].}
}



\maketitle

\begin{abstract}
As seasonal thermal energy storage emerges as an efficient solution to reduce CO$\mathbf{_2}$ emissions of buildings, challenges appear related to its optimal operation.
In a system including short-term electricity storage, long-term heat storage, and where electricity and heat networks are connected through a heat pump, it becomes crucial to operate the system on two time scales.
Based on real data from a university building, we simulate the operation of such a system over a year, comparing different strategies based on model predictive control (MPC).
The first objective of this paper is to determine the minimum prediction horizon to retrieve the results of the full-horizon operation problem with cost minimization.
The second objective is to evaluate a method that combines MPC with setting targets on the heat storage level at the end of the prediction horizon, based on historical data. For a prediction horizon of 6 days, the suboptimality gap with the full-horizon results is 4.31\%, compared to 11.42\% when using a prediction horizon of 42 days and fixing the final level to be equal to the initial level, which is a common approach.
\end{abstract}


\section{Introduction}

As buildings are one of the main contributors of CO$\mathrm{_2}$ emissions worldwide, a shift to more sustainable buildings is necessary to mitigate climate change.
One way to achieve this is by enhancing existing buildings to render them more self-sufficient and to prioritize sources of energy with a low environmental impact.
An example of such a project is underway for a building on the campus of Ecole Polytechnique, in France. It will be equipped with solar photovoltaics (PV) and electricity storage on the electricity side, and solar thermal and seasonal heat storage on the heat side, and both systems will be connected through a heat pump.
Seasonal thermal storage is a solution that can further accelerate the decarbonization of buildings compared to shorter-term thermal storage systems~\cite{noauthor_stockage_2023}. If optimizing the daily operation of such a system, but with a myopic view, the importance of the potential charge of the seasonal heat storage system over the summer might be overlooked.
On the other hand, considering long horizons can result in excessive running times and inadequate solutions, as forecast accuracy degrades with longer forecast horizons.


There is a large literature on the operation of similar systems. 
Examples include \cite{lu_optimal_2015, kriett_optimal_2012, zhang_uncertainty-resistant_2019, farmani_conceptual_2018}, which all feature thermal storage, but with capacities of a couple of days at most, such that the question of seasonal storage does not arise.
Fewer works do consider long-term or seasonal storage \cite{weber_model_2022, thaler_hybrid_2023, castelli_robust_2023}. 

Such operation problems are usually solved with model predictive control, which uses a rolling horizon, or receding horizon \cite{kriett_optimal_2012,  zhang_uncertainty-resistant_2019}. To make the decisions for the next period starting at time $t$, the \emph{control horizon} $H$, the problem is solved over a longer horizon, the \emph{prediction horizon} $T$. This is then repeated at time $t+H$.
The choice of the prediction horizon is usually arbitrary, though it is known that it can have a considerable impact on the quality of the results \cite{parisio_use_2014}.
An exception is~\cite{cruise_control_2019}, in which a minimum prediction horizon is obtained for a storage scheduling problem.
A \emph{minimum prediction horizon} is a prediction horizon for which it is guaranteed that the decisions over the control horizon match the decisions in the complete problem, which could be a finite-horizon problem used for benchmarking purposes. We refer to the complete problem as the \emph{full-horizon} problem. 
Their approach does not apply to systems with more components~\cite{cruise_control_2019}.

Current solutions to schedule long-term storage include using longer prediction horizons, possibly with a coarser resolution \cite{cuisinier_new_2022}, and including end-of-horizon targets \cite{weber_model_2022, castelli_robust_2023, guerra_towards_2024} or a value for the stored energy \cite{weber_model_2022, thaler_hybrid_2023, guerra_towards_2024}. A value for the stored energy is hard to determine and impacts the objective function, while a target level is more tangible. However, even with a target level, the choice of the prediction horizon remains a pressing open question, particularly because it might lead to infeasibilities if chosen too short.

\begin{figure*}[tb]
  \centering
    {\includegraphics[width=0.775\linewidth]{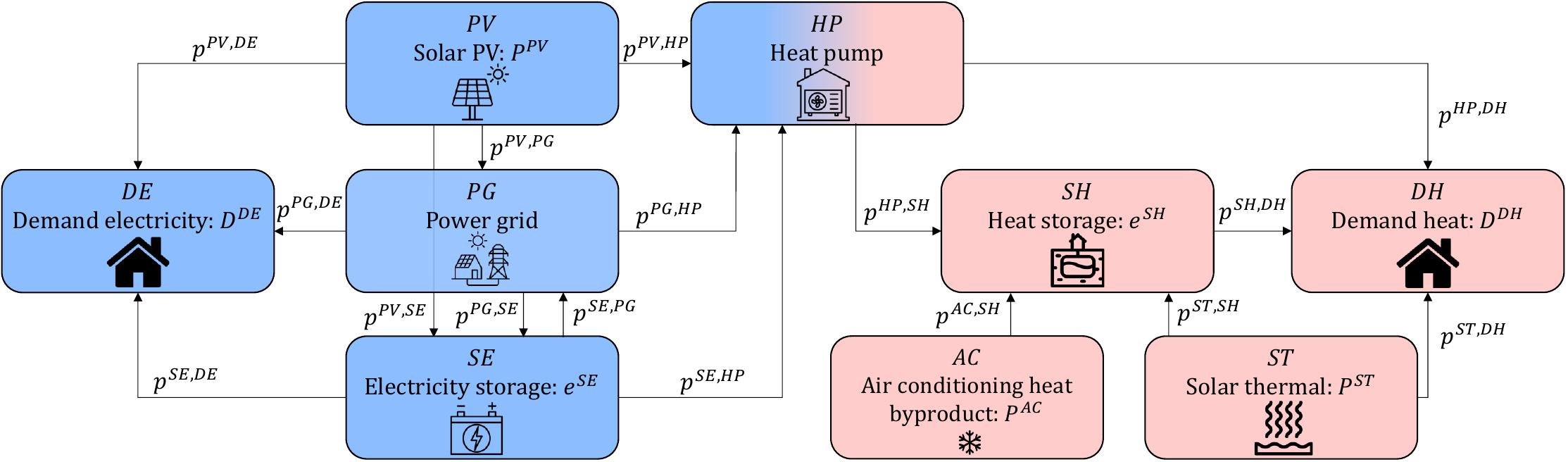}}
  \caption{Representation of the system to operate}
  \label{fig:System}
\end{figure*}

Based on the operation of a realistic building, we make the following contributions. First, we study the minimum length of the prediction horizon.
We then evaluate an alternative approach, using target levels for the long-term thermal storage system at the end of the prediction horizon based on historical data. We show the advantage of this approach by comparing the operation of the system over a year to using a longer prediction horizon but letting the final level return to its initial value, which is a common approach. We show in the example of the method with historical targets that when the prediction horizon is too short, the problem may become infeasible.
\IEEEpubidadjcol

In Section \ref{sec:system}, we present the model and the main assumptions. Section \ref{sec:case} introduces the case study and the full-horizon benchmark. In Section \ref{sec:opt}, we evaluate the minimum prediction horizon, propose an alternative approach, and assess its performance. We present conclusions in Section~\ref{sec:ccl}.

\section{System and Model} \label{sec:system}

We consider the problem of scheduling the energy transfers for the building of Fig. \ref{fig:System}, 
with a focus on the operation of the 
storage systems. This system can also produce electricity and heat for its own consumption, using solar devices. The excess electricity can be injected into the power grid. 

\subsection{Optimization Model}

We want to optimize the operation of the system given in Fig.~\ref{fig:System} over time horizon $\mathcal{T} = \{1,..,T\}$, with resolution $\Delta t$.

\subsubsection{Network Representation of the System}

We consider the components of the system as nodes, gathered in $\mathcal{N} = \{ DE,$ $PV, PG, SE, HP, SH, AC, ST, DH\}$, where the abbreviations for the different components are as shown in Fig.\ref{fig:System}.
Subset $\mathcal{N}^\text{f}_n$, with $n \in \mathcal{N}$ gathers the nodes with a flow to node $n$.
Similarly, $\mathcal{N}^\text{t}_n$ gathers the nodes to which node $n$ can have a flow.
The definition of these subsets corresponds to the arrows in Fig. \ref{fig:System}. For example, $\mathcal{N}^\text{f}_{PG}= \{PV, SE\}$.
The principal decision variables are $p^{n_1,n_2}_t$, which represents the power flow from $n_1$ to $n_2$ for time $t$.

\subsubsection{Constraints per Component}

\paragraph{Demands}
For electrical and heat loads, the demands $D_{t}^{DE}$ and $D_{t}^{DH}$ have to be met: 
\begin{align}
    \label{eq:dmd} D_{t}^{n} = \sum_{n_1\in \mathcal{N}^\text{f}_n} p^{n_1,n}_t, &&& \forall t \in \mathcal{T}, \,  n \in \{DE,DH\}.
\end{align}

\paragraph{Solar PV}
The electricity produced by the solar PV, $P^{PV}_t$, is used to fulfill the demand, to charge the storage system, by the heat pump, and/or to be injected into the grid:
\begin{align}
    \label{eq:pv} P_{t}^{PV} = \sum_{n_2\in \mathcal{N}^\text{t}_{PV}} p^{PV,n_2}_t , &&& \forall t \in \mathcal{T}.
\end{align}

\paragraph{Solar Thermal}
The production of the solar thermal system, $P_{t}^{ST}$, can be used to fulfill the demand and/or to charge the storage system, or it can be lost if in excess:
\begin{align}
    \label{eq:st} P_{t}^{ST} \geq \sum_{n_2\in \mathcal{N}^\text{t}_{ST}} p^{ST,n_2}_t , &&& \forall t \in \mathcal{T}.
\end{align}

\paragraph{Air Conditioning}
The heat that can be stored from AC in summer is bounded by the total heat byproduct, $P^{AC}_t$:
\begin{align}
    \label{eq:ac} P_{t}^{AC} \geq \sum_{n_2\in \mathcal{N}^\text{t}_{AC}} p^{AC,n_2}_t , &&& \forall t \in \mathcal{T}.
\end{align}

\paragraph{Storage Systems}
The two storage systems are modeled similarly. We introduce the variables $e_{t}^{SE}$ and $e_{t}^{SH}$ to monitor the state of energy of the electricity storage and of the heat storage, respectively. The state of energy is updated by considering charged and discharged quantities, where some energy is lost due to inefficiencies when charging, $\eta^{SE,\text{ch}}$ and $\eta^{SH,\text{ch}}$, and when discharging, $\eta^{SE,\text{dis}}$ and $\eta^{SH,\text{dis}}$. Moreover, stored energy is also assumed to leak.
The proportion that remains is identified by $\rho^{SE}$ and $\rho^{SH}$, respectively, where $0<\rho^n\leq 1$. Altogether the state of energy follows
\begin{align}
     \nonumber e^n_{t} = \rho^n e^n_{t-1} + \Delta t & \left ( \eta^{n,\text{ch}} \sum_{n_1 \in \mathcal{N}^\text{f}_n}  p_t^{n_1,n} - \frac{1}{\eta^{n,\text{dis}}} \sum_{n_2 \in \mathcal{N}^\text{t}_n} p_t^{n,n_2} \right ),\\
    \label{eq:soe} & \quad \quad \forall t \in \mathcal{T} \setminus \{1\}, \, n \in \{SE,SH\}.
\end{align}
For the first time period, we have to consider the initial state of energy $E^{SE,\text{init}}$ for the electricity storage and $E^{SH,\text{init}}$ for the heat storage. This gives
\begin{align}
    \nonumber  e^n_{1} = \rho^n E^{n,\text{init}} + \Delta t & \left ( \eta^{n,\text{ch}} \sum_{n_1 \in \mathcal{N}^\text{f}_n} p_1^{n_1,n} - \frac{1}{\eta^{n,\text{dis}}} \sum_{n_2 \in \mathcal{N}^\text{t}_n} p_1^{n,n_2} \right ), \\
    \label{eq:soe_init} & \quad \qquad \qquad \qquad \forall n \in \{SE,SH\}.
\end{align}
We also include a constraint to set the state of energy at the end of the horizon to pre-defined values, $E^{SE,\text{end}}$ and $E^{SH,\text{end}}$:
\begin{align}
    \label{eq:soe_end}  e^n_{T} = E^{n,\text{end}}, &&& \forall n \in \{SE,SH\} .
\end{align}
The state of energy is further limited by lower bounds, $\underline{E}^{SE}$ and $\underline{E}^{SH}$, and upper bounds $\overline{E}^{SE}$ and $\overline{E}^{SH}$:
\begin{align}
    \label{eq:soe_bounds} \underline{E}^{n} \leq e^n_{1} \leq \overline{E}^{n}, &&& \forall t \in \mathcal{T}, \, n \in \{SE,SH\}.
\end{align}
Further, there are bounds on the power that can be charged, $\bar{P}^{SE,\text{ch}}$ and $\bar{P}^{SH,\text{ch}}$.
The binary variables $y_t^{SE}$ and $y_t^{SH}$ are introduced to choose between charge and discharge.
A value of~1 indicates that the corresponding storage system is charging, while a value of 0 indicates that it is discharging. This gives
\begin{align}
    \label{eq:ch_bounds} \sum_{n_1 \in \mathcal{N}^\text{f}_n} p_t^{n_1,n} \leq y_t^{n} & \bar{P}^{n,\text{ch}}, &&& \forall t \in \mathcal{T}, \, n \in \{SE,SH\}.
\end{align}
The power discharged has upper bounds $\bar{P}^{SE,\text{dis}}$ and $\bar{P}^{SH,\text{dis}}$:
\begin{align} 
    \nonumber \sum_{n_2 \in \mathcal{N}^\text{t}_n} p_t^{n,n_2} \leq (1-y_t^{n}) & \bar{P}^{n,\text{dis}}, \\
    \label{eq:dis_bounds} & \forall t \in \mathcal{T}, \, n \in \{SE,SH\}.
\end{align}

\paragraph{Heat Pump}
The heat pump converts electricity to heat with a coefficient of performance $COP$:
\begin{align}
     \label{eq:hp} \sum_{n_2\in \mathcal{N}^\text{t}_{HP}} p^{HP,n_2}_t = COP \sum_{n_1\in \mathcal{N}^\text{f}_{HP}} & p^{n_1,HP}_t , &&& \forall t \in \mathcal{T}.
\end{align}
It also has a capacity $\bar{P}^{HP}$ for heat production:
\begin{align}
    \label{eq:hp_bound} \sum_{n_2\in \mathcal{N}^\text{t}_{HP}} p^{HP,n_2}_t  \leq \bar{P}^{HP}, &&& \forall t \in \mathcal{T}.
\end{align}

\subsubsection{Objective}
We consider the control of the system to minimize the operating costs. Costs at time $t$ are given by the difference between the costs of using electricity from the grid, with price $C_{t}\uptxt{buy}$, and the revenues from injecting power into the grid, with price $C_{t}\uptxt{sell}$. The objective function is then
\begin{align} \label{eq:obj}
    \nonumber F(\mathbf{p}) = & \sum_{t \in \mathcal{T}} \Delta t \left  (C_{t}\uptxt{buy} \sum_{n_2 \in \mathcal{N}^\text{t}_{PG}} p^{PG,n_2}   \right.\\
    & \qquad \qquad \qquad \qquad \left. - C_{t}\uptxt{sell} \sum_{n_1\in \mathcal{N}^\text{f}_{PG}} p^{n_1,PG}_t  \right ),
\end{align}
where $\mathbf{p}$ gathers all the flow variables and $\mathbf{e}$ gathers all the state of energy variables.

\subsubsection{Model}
The resulting optimization model is:
\begin{subequations} \label{pb:opt}
\allowdisplaybreaks
    \begin{align}
        \label{eq:min} \min_\mathbf{p, e, y} \quad & F(\mathbf{p}) \\
        \label{eq:cstr}\text{s.t.}\quad & \eqref{eq:dmd} - \eqref{eq:hp_bound}\\
        \label{eq:bin} & y_t^{n} \in \{0,1\}, &&& \forall t \in \mathcal{T}, \, n \in \{SE,SH\},\\      
        \label{eq:pos} & p_t^{n_1,n_2} \geq 0, &&& \forall t \in \mathcal{T}, \, n_1 \in \mathcal{N}, \, n_2 \in \mathcal{N},
    \end{align}
\end{subequations}
where $\mathbf{y}$ gathers all the binary variables.

\subsubsection{Relaxation}

Under certain conditions, solutions with simultaneous charge and discharge are suboptimal. Here, we observe that simultaneous charge and discharge can only be profitable when the prices are negative, so we remove the binary variables for the hours with positive prices. We drop the corresponding time indices in \eqref{eq:bin}, and in \eqref{eq:ch_bounds} and \eqref{eq:dis_bounds} we replace the corresponding binary variables with 1.

\subsection{Assumptions}
On the heat side, we consider the exchanges in terms of power. Modeling with temperature would allow a more accurate representation of the heat exchanges but it would make the problem significantly more complex to solve.
We consider that the capabilities of the storage systems stay the same over the considered horizon. They can charge or discharge at any power between the bounds, meaning that we assume continuity for the control of storage systems.
It is always possible to sell electricity to the market. The system is too small to have an impact on the market prices.

\section{Case Study} \label{sec:case}
We evaluate the operation of the system of Fig. \ref{fig:System} over a year, with an hourly resolution, using data from a university campus building, between 2020 and 2022. The code and data are available online ({https://github.com/eleaprat/residential-SG}). We assume perfect foresight of future values.

\subsection{Data}

\paragraph{Electrical Load}
For the demand for electricity, we use the hourly consumption of the building.

\begin{figure}[bt]
    \centering
    \resizebox{0.98\linewidth}{!}{
    \includegraphics{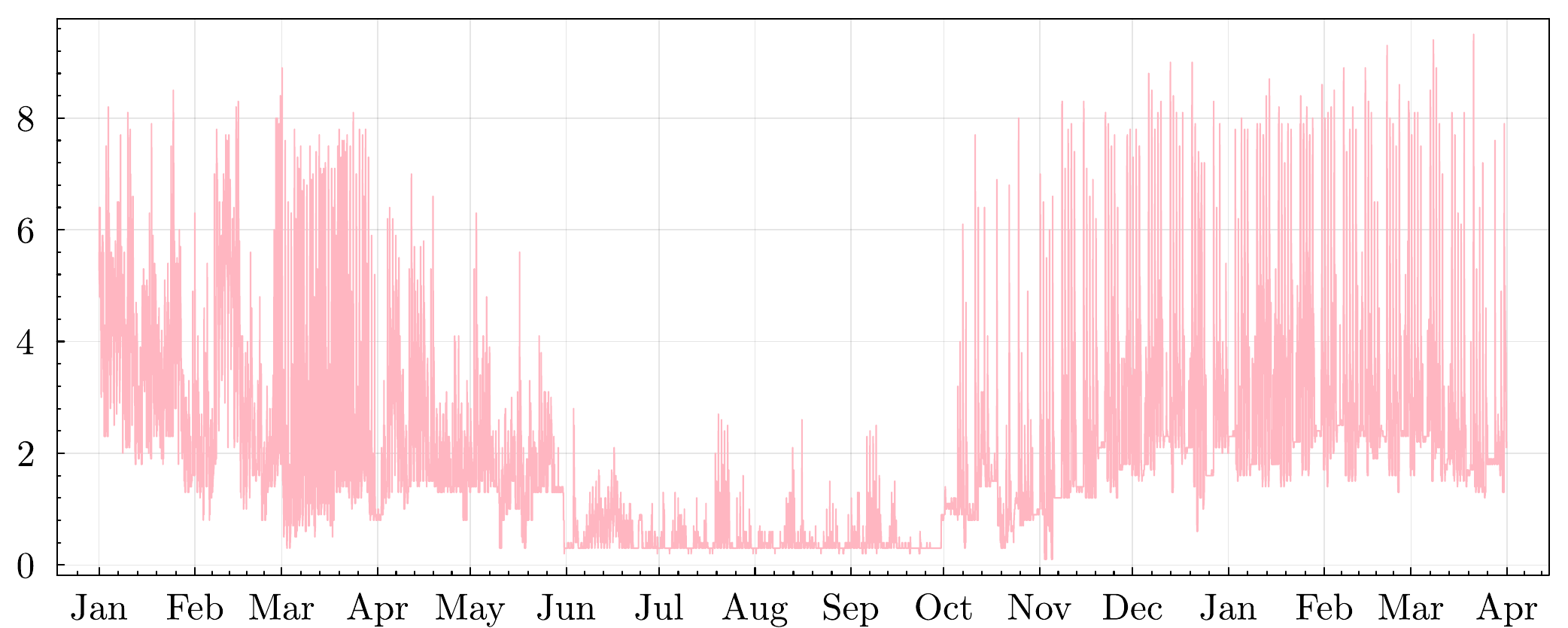}}
    \caption{Electric heat consumption including AC (kWh) for 2021 and beginning of 2022}
    \label{fig:heat}
\end{figure}

\paragraph{Heat Load and Air Conditioning}
Regarding the demand for heating, we use the total consumption of electricity for heating, as shown in Fig. \ref{fig:heat}. We assume that the consumption of electricity by the heaters corresponds to the heat demand in winter and to the heat production resulting from the use of the AC in summer, from the $1\uptxt{st}$ of June to the $30\uptxt{th}$ of September. This is an approximation; in practice, we expect AC and heating to overlap for a part of the year. However, we can observe in Fig. \ref{fig:heat} that the consumption pattern noticeably changes in this summer period.

\paragraph{Solar PV}
We use data for the hourly production of one panel with a nominal power of 250 W installed on the same building. We assume that 80 panels are installed.

\paragraph{Solar Thermal}
We use the irradiance data. We multiply it by an efficiency of 0.9 to get the production for 1 m$^2$ of panels. We consider 12 m$^2$ of panels.

\paragraph{Electricity Storage}
The maximum discharge is set to 10 kW and the maximum charge to 16 kW. We set the capacity of the storage system to 49 kWh which is characteristic of a battery. The charge efficiency and the discharge efficiency are 97\% each. The rate of self-discharge is 0.01\% per hour. We impose that the storage system is empty at the beginning and at the end of the year.

\paragraph{Heat Storage}
The maximum discharge is set to 9.18~kW and the maximum charge to 10.2 kW. We set the capacity of the storage system to 4640 kWh.
The charge efficiency and the discharge efficiency are 78\% each.
The rate of self-discharge is 0.007\% per hour.
We set the initial level at the beginning of the year and the final level at the end of the year to 3000 kWh.

Note that the targets for both storage systems are only introduced to facilitate comparison between different approaches over a year. In a real application, the transition between two years would be continuous.

\paragraph{Power Grid}
We use the hourly spot electricity price for the zone DK2 of Denmark. When buying electricity, a fixed transport fee of 0.20€/kWh is added.

\paragraph{Heat Pump}
The capacity for heat production is 15~kW and the coefficient of performance is 4.

\subsection{Full-Horizon Benchmark}
\begin{figure}[tb]
    \centering
    \subfloat[\label{fig:res_fh}]
    {\includegraphics[trim={0 0 -1cm 0},clip,width=0.87\linewidth]{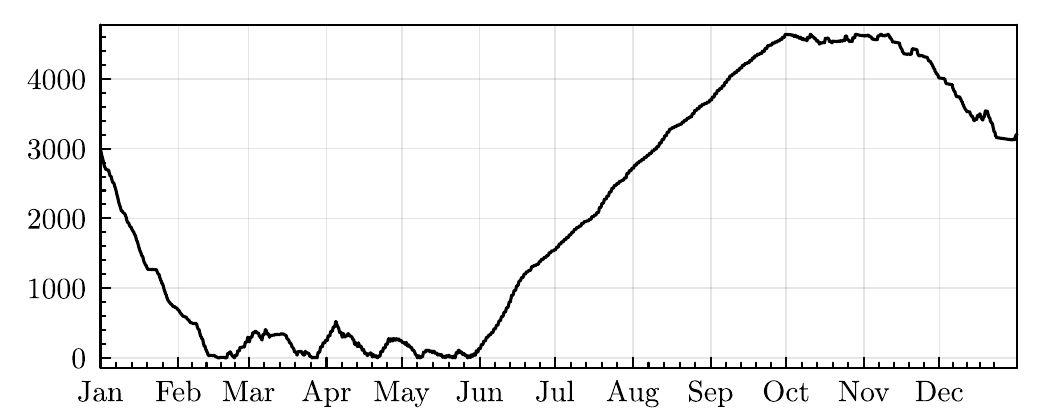}}
    \\
    \subfloat[\label{fig:min_horizon_both}]
    {\includegraphics[trim={0 0 1.7cm 0},clip,width=0.87\linewidth,]{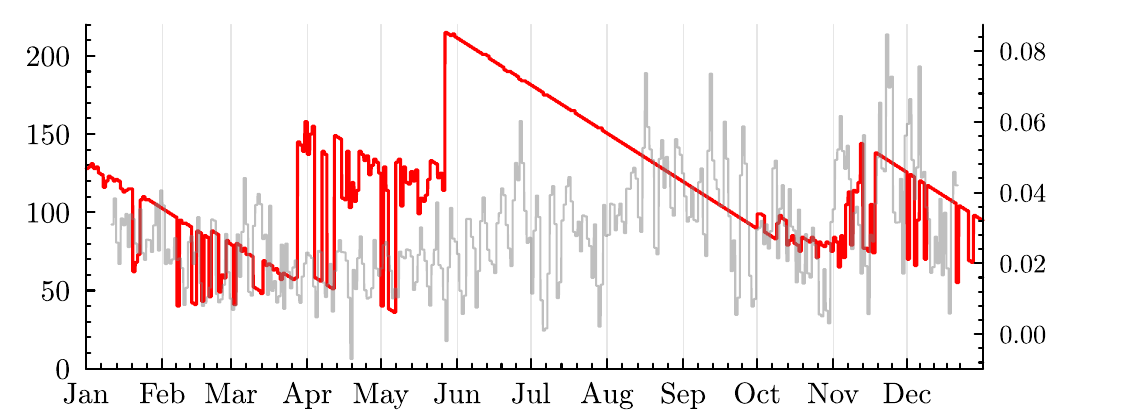}} 
    \caption{Plots for 2021 of (a) the state of energy for the heat storage for the full-horizon benchmark (kWh), (b) the minimum number of days in the prediction horizon to recover the results of the full-horizon problem on the left axis (red), and on the right axis the average daily electricity price (€/kWh).}
    \label{fig:plots}
\end{figure}

We optimize the system for a full year at once to give us an ideal schedule to compare to. 
The resulting evolution of the state of energy for the heat storage system is shown in Fig.~\ref{fig:res_fh}. The seasonal character of optimal operation of the thermal storage system is evident, where we can see that the storage system is empty at the end of February and exclusively charges between the end of May and the beginning of October. This is the behavior that we need to preserve in the optimization of the daily operation of the system.

\section{Optimization with Storage Systems} \label{sec:opt}

Every day, the schedule for the operation of the system should be decided. However, the longer term should not be disregarded, at the risk of making myopic decisions.
To avoid this, we consider the use of a rolling-horizon approach.

\subsection{Evaluation of the Minimum Prediction Horizon} \label{sec:hor}

For each day of the year considered, we evaluate the minimum prediction horizon needed to obtain solutions that are optimal for the full-horizon problem. We achieve this by running the problem with a prediction horizon of increasing length.
For this study, rather than imposing a level at the end of the year, we use the data at the beginning of 2022. In this way, the results at the end of the year are not impacted by the target.
We run two versions of the problem. In the first one, we force the storage systems to reach the minimum level at the end of the prediction horizon, and in the second one, we force them to reach the maximum level at the end of the prediction horizon.
If the states of energy at the end of the first day for both approaches are equal, then the prediction horizon is long enough: no matter the forecast after the end of the prediction horizon, the decision for the next day will not change. If they are different, we increase the prediction horizon by one day.
We obtain the plot of Fig.~\ref{fig:min_horizon_both}, which shows the minimum length of the prediction horizon in the number of days, for each day of the year. 


We observe that the minimum length of the prediction horizon varies from day to day. 
The minimum prediction horizon for the whole year is 36 days. We can compare this to the duration of charge, $\Delta t^{n,\text{ch}} = \frac{\overline{E}^n - \underline{E}^n}{\eta^{n,\text{ch}} \overline{P}^{n,\text{ch}}}$, and to the duration of discharge, $\Delta t^{n,\text{dis}} = \frac{\overline{E}^n - \underline{E}^n}{\frac{1}{\eta^{n,\text{dis}}} \overline{P}^{n,\text{dis}}}$, for both storage systems. For the electricity storage system, $\Delta t^{SE,\text{ch}} = 3.16$ hours and $\Delta t^{SE,\text{dis}} = 4.75$ hours, and for the thermal storage system, $\Delta t^{SH,\text{ch}} = 583.2$ hours and $\Delta t^{SH,\text{dis}} = 394.3$ hours, which corresponds to more than 24 and 16 days, respectively. This suggests that the prediction horizon has to be at least long enough for the thermal storage to fully charge.
To ensure that the thermal storage system charges during summer, the horizon must incorporate the forecast demand for heat in the coming winter, which explains why the minimum prediction horizon is much longer.
There is a pattern of a general decrease of the minimum length of the prediction horizon followed by a significant increase. Looking in parallel at Fig. \ref{fig:res_fh}, we see that these significant increases coincide with the storage reaching a bound.
Other factors affect the length of the minimum prediction horizon and in particular the variability of the prices, which can also be observed on Fig. \ref{fig:min_horizon_both}.

\subsection{Hybrid Approach} \label{sec:hybrid}
Since the thermal storage system is driving the minimum length of the prediction horizon, we consider a hybrid approach, where we use a prediction horizon of 4 days and include a hard constraint on the final state of energy of the heat storage. This end-of-horizon value serves as a guide and does not necessarily correspond to an actual command for that day since only the results of the first day are implemented. 
A question is then how to evaluate this target level. In this study, for a simple proof of concept, we use the optimal levels for the year 2020, which are obtained by optimizing the operation of the complete system over the whole year of 2020 at once, with the same initial and final levels.
For the electricity storage system on the other hand, 4 days is sufficient to capture future opportunities and the final level at the end of these 4 days can be left free without impacting the quality of the solution.

\subsection{Comparison of Different Approaches}
\begin{figure}[tbh]
    \centering
    \includegraphics[width=0.99\linewidth]{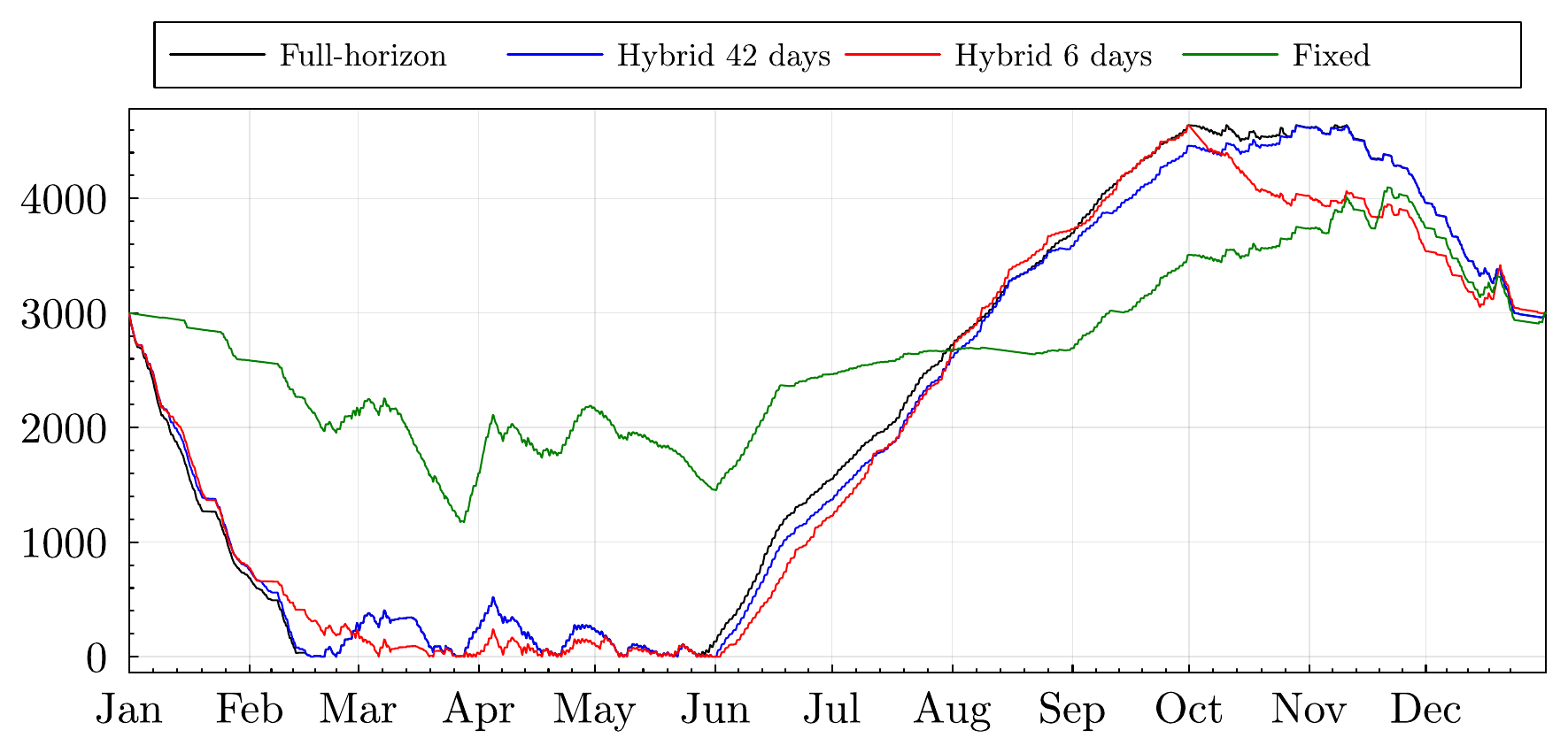}
    \caption{State of energy for the thermal storage (kWh) for 2021, for the different methods} \label{fig:res_all_soe}
\end{figure}

We compare the results from the hybrid approach to the full-horizon benchmark, and to the most common approach in the literature, which is to consider a rolling horizon with a prediction horizon sufficiently long to allow for the full charge and discharge of the storage systems, 42 days in our case\footnote{For the heat storage, when considering self-discharge.}, and with the final level of the storage systems equal to their initial level.
For the hybrid approach, we first consider a prediction horizon of 4 days and increase the length up to 42 days, to match the rolling horizon with fixed final level.
The results in terms of costs for the year, and suboptimality of the objective value compared to the full-horizon approach are given in Table~\ref{tab:res}.
We observe that for prediction horizons of less than 6 days, the problem is infeasible.
Indeed, when we impose a varying final level on the thermal storage system, the prediction horizon needs to be long enough to ensure that this level can be reached.
Note that this could also be resolved by introducing slack variables in~\eqref{eq:soe_end}, with sufficiently high penalties in the objective function \cite{castelli_robust_2023}.
The maximum gap with the hybrid approach is 4.31\% and it reduces to 0.92\% when a prediction horizon of 42 days is used, compared to 11.42\% for the rolling horizon with fixed final level.
We can also see the impact on runtime.
The state of energy of the thermal storage system is indicated in Fig.~\ref{fig:res_all_soe}. For the hybrid approach, the results for a prediction horizon of 6 days and for a prediction horizon of 42 days are plotted.
The capability of the combined approach to consider the longer term, while still being flexible to short-term realizations is clearly visible. On the other hand, the approach with fixed level only follows the short-term variations, while long-term patterns are lost.

\begin{table}[tb]
\caption{Results for the operation over 2021}
\centering
\resizebox{0.99\linewidth}{!}{%
\begin{tabular}{lrrrr}
\hline
\textbf{Method} & \multicolumn{1}{l}{Horizon (days)} & \multicolumn{1}{l}{Costs (€)} & \multicolumn{1}{l}{Subopt. gap} & \multicolumn{1}{l}{Runtime (s)} \\ \hline
Full-horizon & 365 & 1362.45 & 0\% & 20\\ \hline
Hybrid       & 4 & Infeas. & Infeas. & Infeas.\\
approach     & 5 & Infeas. & Infeas. & Infeas.\\
             & 6 & 1421.20 & 4.31\% & 12\\
             & 10 & 1401.55 & 2.87\% & 26\\
             & 20 & 1388.98 & 1.95\% & 54\\
             & 30 & 1382.05 & 1.44\% & 122\\
             & 42 & 1374.95 & 0.92\% & 230 \\ \hline
Fixed level & 42 & 1518.04 & 11.42\% & 177 \\ \hline
\end{tabular}
}%
    \label{tab:res}
\end{table}

\section{Conclusion} \label{sec:ccl}
In this paper, we studied the optimal operation of electricity and heat networks for a building, in the presence of short-term electricity storage and long-term thermal storage. 
While a rolling-horizon approach can help to have a longer-term perspective, the length of the prediction horizon for which the optimal solution of the full-horizon problem can be recovered is more than the duration for which we can obtain reliable forecasts.
We then evaluated a method that combines a rolling horizon with setting a smart value for the long-term storage level at the end of the prediction horizon. We showed that this approach can perform comparatively well when the level is set to the optimal value obtained with the data of the previous year. For a prediction horizon of 6 days, the suboptimality gap with the full-horizon results is 4.31\%, compared to 11.42\% when using a prediction horizon of 42 days and fixing the final level to be equal to the initial level. 
Future work should focus on introducing a more complex version of the system by relaxing some of the assumptions that were made. In particular, it would be relevant to consider an augmented model with temperature and degradation of the storage systems.

\bibliographystyle{IEEEtran}
\bibliography{ref}

\end{document}